\documentclass[a4paper,fleqn]{cas-dc}

\usepackage[authoryear,longnamesfirst]{natbib}

\usepackage{amsfonts}
\usepackage{amsmath}
\usepackage{amssymb}
\usepackage{physics}
\usepackage{xcolor}
\usepackage{float}
\usepackage{CJKutf8}
\usepackage{threeparttable}

\IfFileExists{xurl.sty}{\usepackage{xurl}}{%
  \makeatletter
  \g@addto@macro\UrlBreaks{\do\a\do\b\do\c\do\d\do\e\do\f\do\g\do\h%
    \do\i\do\j\do\k\do\l\do\m\do\n\do\o\do\p\do\q\do\r\do\s%
    \do\t\do\u\do\v\do\w\do\x\do\y\do\z\do\A\do\B\do\C\do\D%
    \do\E\do\F\do\G\do\H\do\I\do\J\do\K\do\L\do\M\do\N\do\O%
    \do\P\do\Q\do\R\do\S\do\T\do\U\do\V\do\W\do\X\do\Y\do\Z%
    \do\0\do\1\do\2\do\3\do\4\do\5\do\6\do\7\do\8\do\9}%
  \makeatother}

\newcommand{\EMT}{\mathcal{T}}
\newcommand{\E}{\mathcal{E}}
\newcommand{\AM}{\mathcal{L}_z}

\newcommand{\lapse}{\mathcal{N}}
\newcommand{\needsnum}[1]{{\color{red}\ifmmode[#1]\else[#1]\fi}}

\newcommand{\zhname}[1]{%
  \ifdefined\XeTeXrevision {\font\zhnamefont="[FandolSong-Regular.otf]" at \csname f@size\endcsname pt \zhnamefont #1}\else\begin{CJK}{UTF8}{gbsn}#1\end{CJK}\fi%
}
\providecommand{\texorpdfstring}[2]{#1}

\graphicspath{{revision_support/}}
\begin{document}
\let\WriteBookmarks\relax

\shorttitle{A Dark Matter Masquerade}
\shortauthors{L. Lui, Z. Zhang \& A. Torres-Orjuela}

\title[mode=title]{A Dark Matter Masquerade: Degeneracies in Black Hole and Accretion Inference from X-ray Reflection Measurements and Prospects for Compact Dark Matter Halo Constraints}

\author[1]{Leif Lui}[orcid=0000-0003-0246-9681]
\fnmark[1]

\author[2,3]{Zijian Zhang}[suffix={(\zhname{张子健})},orcid=0000-0001-8701-2116]
\fnmark[1]
\ead{astrozzj@connect.hku.hk}

\author[1]{Alejandro Torres-Orjuela}[orcid=0000-0002-5467-3505]
\ead{atorreso@bimsa.cn}

\affiliation[1]{organization={Beijing Institute of Mathematical Sciences and Applications}, city={Beijing}, postcode={101408}, country={China}}

\affiliation[2]{organization={Department of Physics, The University of Hong Kong}, addressline={Pokfulam Road}, city={Hong Kong}, country={China}}

\affiliation[3]{organization={The Hong Kong Institute for Astronomy and Astrophysics, The University of Hong Kong}, addressline={Pokfulam Road}, city={Hong Kong}, country={China}}

\fntext[1]{These authors contributed equally to this work.}


\begin{abstract}
X-ray reflection spectroscopy is an established probe of black hole (BH) spin and accretion geometry, and many studies have examined its systematic uncertainties. In this work, we explore a possible environmental effect associated with compact dark matter (DM) halos. We adopt numerical BH-DM spacetimes and perform general-relativistic ray-tracing calculations to generate broadened Fe\,K$\alpha$ line profiles. We find an overall redshift of the line profiles for the configurations examined. We fit these profiles with standard Kerr models and extend the analysis to simulated NICER and NuSTAR spectra by adding a continuum component and accounting for absorption and instrumental effects. The Kerr fits can reproduce the broad spectral shape with different inferred spins, inclinations, and emissivity profiles, revealing degeneracies between halo effects and these source parameters. In several configurations, spectra from slowly spinning BHs can be fitted with high-spin Kerr models. These results suggest that compact DM halos may introduce an additional uncertainty in reflection-based measurements. Conversely, reflection models that include the environment could ultimately help constrain key halo parameters, such as its mass and characteristic scale, complementing gravitational-wave probes.
\end{abstract}


\maketitle

\section{Introduction}
X-ray reflection features are commonly observed in the spectra of active galactic nuclei and black hole (BH) X-ray binaries. Hard X-rays produced by a compact region known as the corona illuminate the accretion disk, where they are reprocessed to form a characteristic reflection spectrum. Because the disk can extend to within a few gravitational radii of the BH, photons originating from its innermost regions experience strong Doppler shifts, gravitational redshift, and light bending. These relativistic effects modify the entire reflection spectrum, broadening and shifting its spectral features. The intrinsically strong Fe K$\alpha$ line is a particularly useful diagnostic. Early studies of this line and the reflection spectrum \citep{Fabian89, George91} laid the foundations for using X-ray reflection spectroscopy to probe spacetime and accretion flows near compact objects. Numerical packages such as \texttt{relxill} \citep{Garcia14} combine local reflection calculations of \texttt{xillver} \citep{Garcia10} with relativistic transfer functions from \citet{Dauser10, Dauser13}, allowing the BH spin, disk inclination, and other accretion parameters to be constrained under the assumption of a vacuum Kerr spacetime. The geometry of the accretion flow can strongly affect the reflection
spectrum. For example, super-Eddington winds can produce features that differ from those of standard thin disks \citep{Thomsen19, Zhang24, Shashank_2025}. Reflection models also make other assumptions, which can introduce systematic uncertainties. The development and the uncertainties are reviewed by, e.g., \citet{Reynolds21, Bambi21, Liu26}. Recently, \citet{Garcia26} synthesized community discussions into a practical framework for assessing reflection-based spin measurements and outlined a roadmap for future analyses in the high-throughput, high-resolution era of X-ray astronomy. In particular, \citet{Bambi17, Abdikamalov19, Abdikamalov20} explore the possibility of testing the Kerr BH hypothesis with X-ray reflections.

Dark matter (DM) around BHs could introduce another systematic uncertainty in X-ray reflection measurements. \citet{Gondolo99} showed that the adiabatic growth of a supermassive BH can steepen the surrounding DM distribution into a dense spike, and \citet{Sadeghian13} extended this scenario into the relativistic regime. Model-dependent evidence for compact DM overdensities around stellar-mass BHs has been inferred from the anomalous orbital decay of black hole X-ray binaries \citep{Chan23}. A similar interpretation has been proposed for the SMBH binary candidate OJ~287, although subsequent studies have questioned whether the observed orbital evolution requires a DM spike \citep{Chan24, Deb25}. More broadly, the profound impact of DM environments on BH geometries and binary dynamics has garnered significant recent attention. For instance, DM halos have been modeled extensively using the Einstein cluster framework to study their influence on the central spacetime, and recent work has begun exploring how these environments alter the dynamics and distinguishability of extreme mass ratio inspirals.~\citep{Cardoso22a, Figueiredo2023, Speeney24, Gliorio2025}.

Recent studies have constructed BH spacetimes with surrounding DM: \citet{Cardoso22a} derived analytical solutions for Schwarzschild BHs surrounded by an anisotropic fluid, and \citet{Fernandes25} developed numerical solutions for spinning BHs with similar DM profiles. Most proposed probes target gravitational-wave signals, BH shadows, or Galactic-center dynamics \citep[see][for a recent review]{Bertone24}. These approaches probe complementary systems and parameter regimes, while X-ray reflection spectroscopy offers another way to study such environments using existing observations. Recent studies have explored phenomenological Fe K$\alpha$ signatures in spherically symmetric fermionic DM ~\citep{Crespi2026}. However, a fully relativistic calculation of the Fe lines in an axisymmetric spacetime with a rotating BH immersed in an anisotropic DM halo has yet to be done and is essential to fully understand the impact of astrophysical environments. 

In this paper, we address the challenge of modeling these complex astrophysical environments by performing a series of general-relativistic ray-tracing simulations. Specifically, we investigate the reflection spectra of a rotating BH immersed in an anisotropic DM halo using the metric described by \citet{Fernandes25}. We explore two key aspects: how the presence of a DM halo affects the line profile, and how this impacts the inference of BH and accretion parameters. The paper is structured as follows. Our model setup and methodology are discussed in Section~\ref{method}. We show the Fe\, K $\alpha$ line spectral features and the analysis in Section~\ref{result}.
We summarize and discuss future directions in Section~\ref{discussion}.

\section{Methodology}\label{method}

\subsection{Solving the Einstein Equations for a Black Hole Submerged in a Dark Matter Halo}\label{Methodology_EFE}

To construct the metric governing a spinning BH embedded in a DM halo, we use the spectral solutions of~\citet{Fernandes25}, who solved the Einstein field equations sourced by the energy-momentum tensor of an anisotropic fluid,
\begin{equation}
\begin{split}
        \EMT_{\mu\nu}=&(\varepsilon+p_1)u_{\mu}u_{\nu}+p_1g_{\mu\nu}\\
        &+(p_r-p_1)k_{\mu}k_{\nu}+(p_2-p_1)s_{\mu}s_{\nu},\\
\end{split}
\end{equation}
where $\varepsilon$, $p_r$, and $p_{1,2}$ are the energy density, radial pressure, and transverse pressures in the fluid's comoving frame, respectively; $u^\mu$ is the fluid four-velocity, and $k^\mu$ and $s^\mu$ are spacelike vectors defining the anisotropy directions. We use the metric signature $(-,+,+,+)$ and geometrized units, where $G=c=1$. The model closes the matter sector by setting $p_r=0$. This corresponds to an Einstein cluster, where a collisionless swarm of particles is supported entirely by tangential stresses with zero radial pressure \cite{Maeda2025, Fernandes25} and by choosing the halo rotation law $\Omega_{\rm halo}=\omega$. The halo therefore has zero velocity relative to a local zero-angular-momentum observer (ZAMO) and carries no angular momentum, although its coordinate angular velocity is generally nonzero. Consequently, the total angular momentum of the spacetime is solely the BH angular momentum, $J=J_{\rm H}$, and the total asymptotic ADM mass satisfies $M_{\rm ADM}=M_{\rm BH}+M_h$. Here, $M_h$ denotes the mass of the DM halo, and $M_{\rm BH} = \kappa/(4\pi)A_H + 2\Omega_H J_H$ is the local horizon mass of the interacting BH (defined exactly by its surface gravity $\kappa$, area $A_H$, and angular velocity $\Omega_H$), rather than the ADM mass of an isolated vacuum BH~\citep{Fernandes25}. We define the dimensionless BH spin as $j=J_{\rm H}/M_{\rm BH}^{2}$, and express all dimensional quantities in units of $M_{\rm BH}$ unless stated otherwise.

We focus on halos with an energy density that follows a Hernquist-like profile~\citep{Hernquist90, Fernandes25},
\begin{equation}\label{eq:hernquist}
    \varepsilon=\frac{M_0 (a_0+r_H)}{2\pi r(r+a_0)^3b^5}\left(1-\frac{r_H}{r}\right)^2,
\end{equation}
where $a_0$ represents the characteristic length scale associated with the halo and $r_H$ is the quasi-isotropic coordinate location of the event horizon. The parameter $M_0$ corresponds to the total halo mass, $M_h$, to leading order in $M_0/a_0$ within the static limit. The function $b(r)>0$ ensures that the density remains non-negative. The function $b(r)>0$ ensures that the density remains non-negative. We use $M_h\simeq10\,M_{\rm BH}$ and $a_0\simeq100\,M_{\rm BH}$, corresponding to a relatively-high compactness $M_h/a_0\simeq0.1$ following ~\cite{Fernandes25}. For the full expression of $b(r)$ and the interpretation of its parameters, see the Supplemental Material of \citet{Fernandes25}. Related properties of BHs surrounded by Hernquist-like or more general matter distributions are discussed by \citet{Cardoso22a, Speeney24}.

For stationary, axisymmetric BHs, the line element in quasi-isotropic coordinates $(t,r,\theta,\phi)$ takes the form
\begin{equation}\label{metric}
\dd s^2=g_{tt}\dd t^2+g_{rr}\dd r^2+g_{\theta\theta}\dd \theta^2+g_{\phi\phi}\dd \phi^2+2g_{t\phi}\dd t\dd\phi,
\end{equation}
where the metric components are functions of $r$ and $\theta$ alone. In terms of the metric functions introduced below, regularity at the horizon requires $\partial_r f=\partial_r g=\partial_r h=\partial_r p_1=\partial_r p_2=0$ and $\omega=\Omega_H=\mathrm{const.}$ at $r=r_H$. Asymptotic flatness requires $f,g,h\to1$ and $\omega,p_1,p_2\to0$ as $r\to\infty$. Axial symmetry and equatorial parity supply the corresponding angular boundary conditions at $\theta=0$ and $\pi/2$~\citep{Fernandes25}. It is convenient to introduce the compactified radial coordinate $x=1-2r_H/r$, where $x\in[-1,1]$.

After imposing these boundary conditions, the Einstein equations can be solved spectrally~\citep{Fernandes23}, and the metric in quasi-isotropic coordinates can be written as
\begin{equation}\label{eq:metric}
\begin{split}
        \dd s^2=-f\frac{N_-^2}{N_+^2}\dd t^2+\frac{g}{f}N_+^4&\left[h\left(\dd r^2+r^2\dd\theta^2\right)\right.\\
        &\quad \left.+r^2\sin^2\theta\left(\dd\phi-\omega\,\dd t\right)^2\right],
\end{split}
\end{equation}
where $N_{\pm}=1\pm r_H/r$ and $\omega$ is the frame-dragging angular velocity. The functions $F^{(i)}=\{f,g,h,\omega\}$ are expanded as
\begin{equation}\label{metric_components}
    F^{(i)}(r,\theta)\approx\sum_{j=0}^{N_x-1}\sum_{k=0}^{N_\theta-1}\alpha_{jk}^{(i)}\,T_j(x)\cos(2k\theta),
\end{equation}
where $N_x$ and $N_\theta$ are the resolutions of the $x$- and $\theta$-grids, $\alpha^{(i)}_{jk}$ are the spectral coefficients, and $T_j(x)$ are Chebyshev polynomials. Spectral methods have been developed for constructing spinning BH solutions in modified gravity \citep{Fernandes23}. Related spectral frameworks for BH perturbations were first developed on a Kerr background in GR \citep{Chung24} and subsequently extended to modified-gravity theories \citep{Chung24a, Chung25}.

\subsection{Geodesic Ray Tracing and Computing the Relativistic X-Ray Reflection Spectrum}
 To account for the light-bending effect and the energy shift induced by the strong gravity field, we perform geodesic ray-tracing with the \texttt{PyHole} package~\citep{Bohn15, Cunha15, Cunha16}, which solves the geodesic equation via Hamilton's equations
\begin{equation}\label{Hamilton}
      \dot{x}^{\mu}=\frac{\partial\mathcal{H}}{\partial p_{\mu}},\quad \dot{p}_{\mu}=-\frac{\partial\mathcal{H}}{\partial x^{\mu}},\quad\mathcal{H}=\frac{1}{2}(g^{\mu\nu}p_{\mu}p_{\nu}+m^2),
\end{equation}
where $m = 0$ is the invariant mass for photons, $x^{\mu}$ and $p_{\mu}$ are the spacetime coordinates and covariant 4-momentum, and the overdot denotes a derivative with respect to an affine parameter. The $t$- and $\phi$-independence of Equation~\eqref{metric} leads to the conserved quantities $p_t=-\E$ and $p_{\phi}=\AM$, which are the photon energy and $z$-angular momentum, respectively. For null geodesics, it is convenient to define the
impact parameter $\lambda = \AM/\E$, which is constant along each ray. 

We approximate an asymptotically distant stationary observer, $u_o^\mu=(1,0,0,0)$, by placing the image plane at $R_{\rm obs}=2000\, M_{\rm BH}$ in our numerical calculations. Furthermore, while we explicitly model the gravitational influence of the DM halo, we treat the accretion disk strictly as a test fluid. The disk's self-gravity is neglected because its mass is typically orders of magnitude smaller than both the central BH and the encompassing halo. Assuming that this accretion disk follows prograde equatorial circular geodesics with angular frequency $\Omega \in \{\Omega_{\mathrm{K}}, \Omega_{\mathrm{DM}}\}$, we write its four-velocity as $u_e^\mu=u_e^t(1,0,0,\Omega)$, where the time component
\begin{equation}\label{eq:ut_2}
u^t_e= \sqrt{\frac{-1}{g_{tt}+2\Omega g_{t\phi}+\Omega^2g_{\phi\phi}}}
\end{equation}
follows directly from the normalization condition $u_\mu u^\mu=-1$. The angular frequency is determined independently by the radial geodesic equation for a circular equatorial orbit
\begin{equation}
    \partial_r g_{tt} + 2\Omega \partial_r g_{t\phi} + \Omega^2 \partial_r g_{\phi\phi} = 0,
\end{equation}
whose prograde root is used in both the Kerr and DM spacetimes~\citep{Bardeen72}. The energy shift between the distant observer and the emitting material on the disk can then be written as
\begin{equation}\label{eq:g_factor}
  g \equiv \frac{E_o}{E_e}
  = \frac{p_\mu u_o^\mu}{p_\mu u_e^\mu}= \frac{-\E}{-\E u^t_e + \AM u^\phi_e} = \frac{1}{u_e^t(1-\lambda\Omega)},
\end{equation}
where $p_\mu$ is the covariant photon 4-momentum obtained from geodesic ray-tracing.
\begin{figure}
    \centering
    \includegraphics[width=0.892\linewidth]{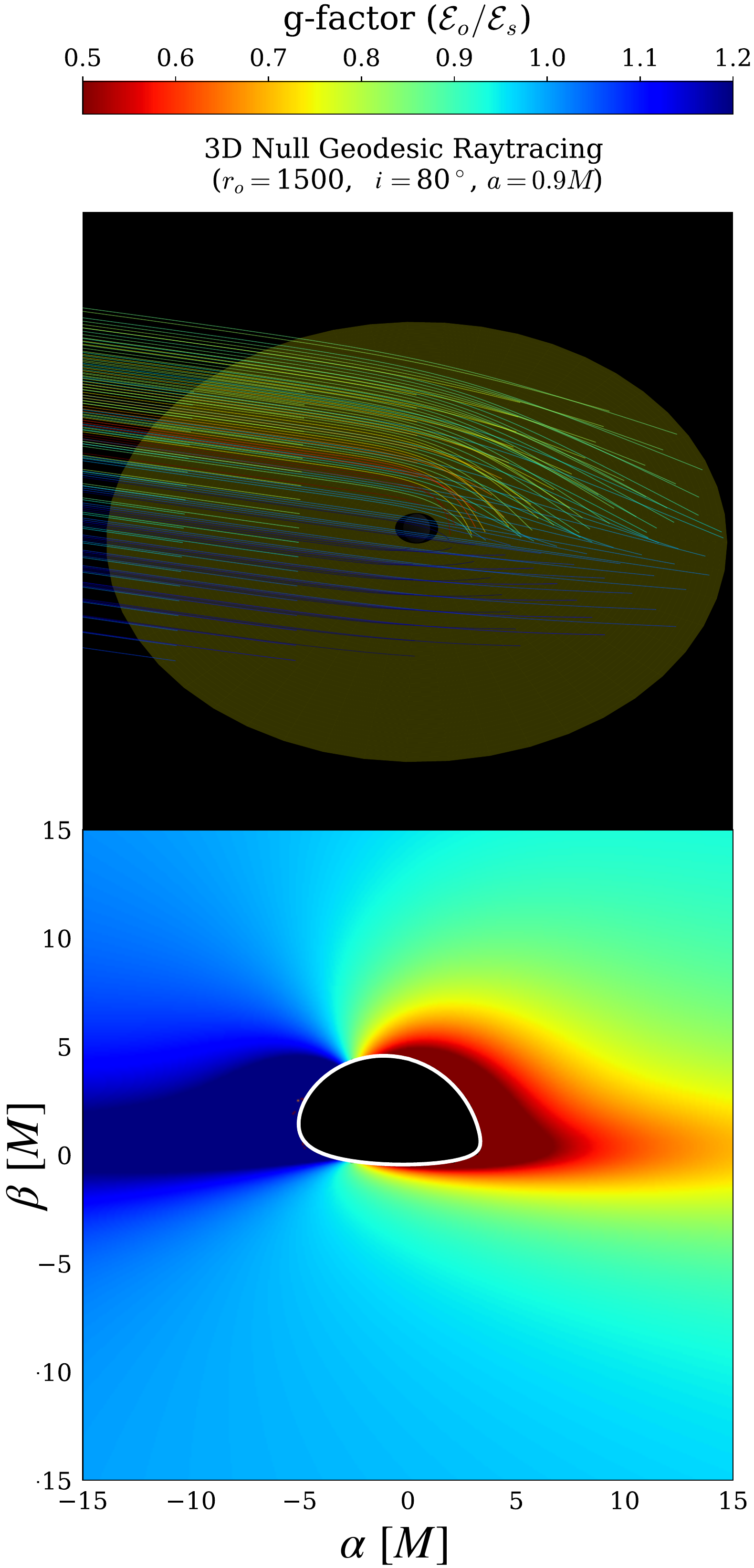}
    \caption{Null geodesic ray-tracing and apparent image of an accretion disk around a Kerr BH ($j=0.9$, $i=80^\circ$). Top panel: 3D null geodesics emitted from the inner region of the disk. Bottom panel: the corresponding 2D apparent image on the observer's sky ($\alpha$, $\beta$), with the projected ISCO outlined in white. The shared colormap denotes the $g$-factor ($E_o/E_e$), highlighting relativistic Doppler beaming on the approaching (left) side and combined kinematic and gravitational redshifting on the receding (right) side.}
    \label{fig:g_factor_plot}
\end{figure}

We perform backwards ray-tracing of parallel photons from the distant observer's image plane $(\alpha,\,\beta)$ until they reach the accretion disk, following \citet{Cunningham73}. The observed energy-specific intensity is related to the emitted intensity through the Lorentz invariance of $I_\nu/\nu^3$ \citep{Lindquist66}, yielding
\begin{equation}
  F_o(E_o)
  = \int
    g^3 I_e\!\left(E_e\right)\dd\alpha\dd\beta,
  \label{eq:flux}
\end{equation}
where $E_e = 6.4$\,keV is the rest-frame energy of the emitting Fe\,K$\alpha$ line. We assume locally isotropic emission of a monochromatic Fe\, K$\alpha$ line. We use a radial power-law emissivity $I_e(r_e)\propto r_e^{-q}$ without introducing a more complex corona setup. We set $q=3$, place the inner disk edge at the prograde ISCO, and use $R_{\rm out}\,=\,100\, M_{\rm BH}$. Rays are traced to their first valid intersection with the optically thick disk, and returning radiation is not included. We show the resulting map of an accretion disk seen by a distant observer in Figure~\ref{fig:g_factor_plot}, recovering the relativistic effects and agreeing qualitatively with previous works~\citep[e.g.,][]{Dauser10b}.

\subsection{Validation}\label{validation}

To validate our implementation, we performed several independent consistency checks. In the Schwarzschild and Kerr limits, as $M_h\to0$, the computed circular-orbit quantities and $g$-factors agree with the corresponding standard Kerr Fe lines from \citet{Dauser10} within the adopted numerical tolerances. Setting the halo density to zero reduces the spectral solution to Kerr and makes the resulting $g_{\rm DM}/g_K$ map consistent with unity. We also compare the Fe K$\alpha$ line profiles from the non-spinning numerical metric with the analytical solution of \citet{Cardoso22a}, and find consistent results. We also recover the ISCO radii tabulated by \citet{Fernandes25}. These tests check the metric interpolation, circular-orbit calculation, and geodesic integration independently.

When radial quantities are compared between the two spacetimes, we match them at fixed circumferential radius $R=\sqrt{g_{\phi\phi}}$, rather than at a fixed quasi-isotropic or Boyer-Lindquist coordinate radius. This is because $R$ represents a physical, gauge-independent proper radius defined by the geometrical norm of the axial Killing vector field.

\section{Results}\label{result}

\subsection{Fe\,K\texorpdfstring{$\alpha$}{alpha} Lines Modified by a Dark Matter Halo}

\begin{figure*}
    \centering
    \includegraphics[width=0.92\linewidth]{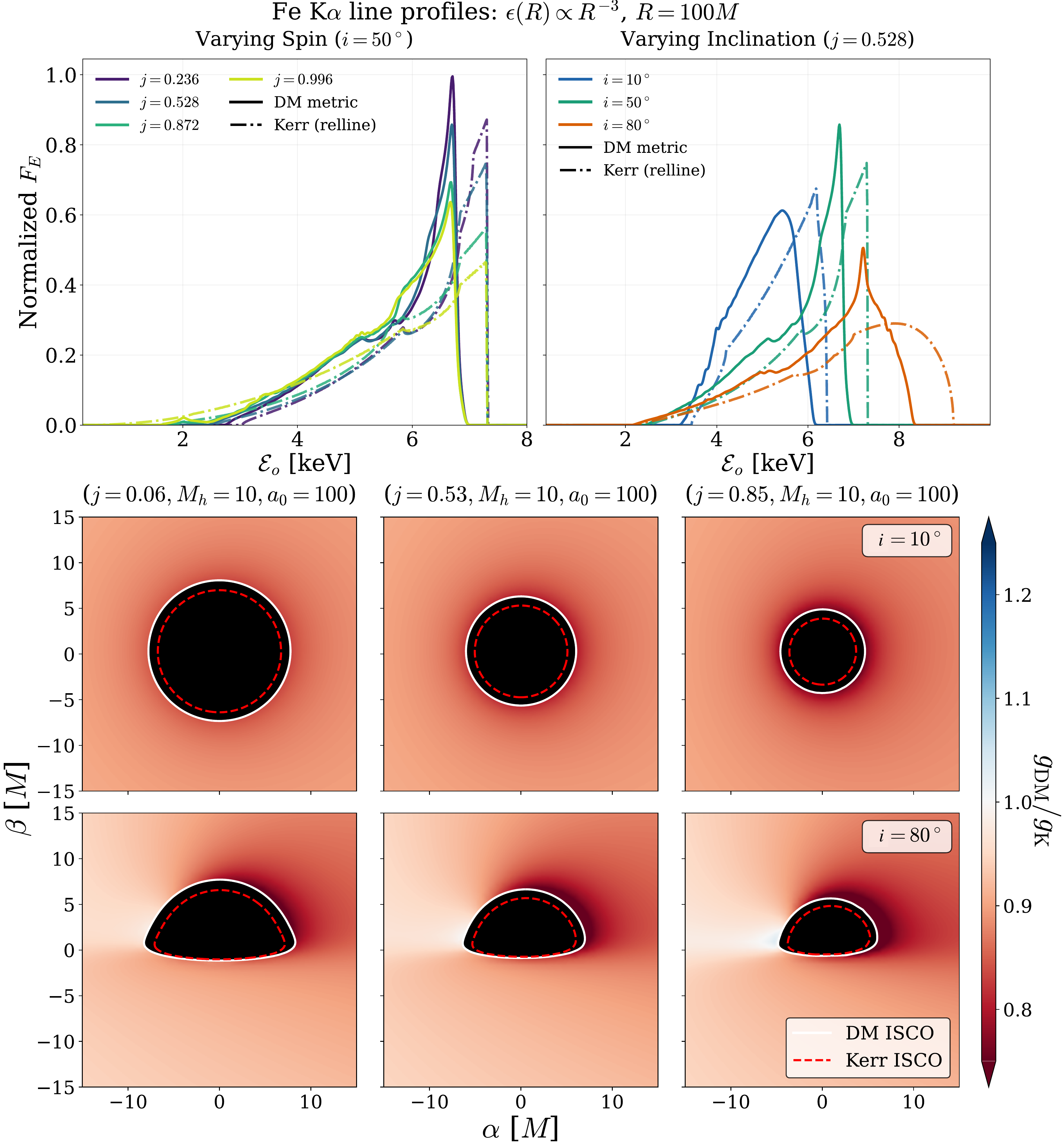}
    \caption{Top: Normalized Fe\,K$\alpha$ line profiles for varying BH spin $j$ at fixed inclination $i=50^\circ$ (left) and for varying inclination at fixed spin $j=0.528$ (right). Profiles in the DM spacetime (solid curves) are compared with their Kerr counterparts with the same $M_{\rm BH}$ and $j$ (dash-dotted curves). The DM profiles are computed for a halo of mass $M_h\simeq10\,M_{\rm BH}$ and characteristic scale $a_0\simeq100\,M_{\rm BH}$, i.e. compactness $M_h/a_0\simeq0.1$. The disk has emissivity $\epsilon(R_{\rm circ})\propto R_{\rm circ}^{-3}$ and $R_{\rm out}=100\,M_{\rm BH}$. Bottom: Ratios $g_{\rm DM}/g_K$ on the observer's image plane $(\alpha,\beta)$ for nearly face-on ($i=10^\circ$, top row) and nearly edge-on ($i=80^\circ$, bottom row) views. Columns show $j\simeq0.06$, $0.53$, and $0.85$ for halo compactness $M_h/a_0\simeq0.1$ ($M_h\simeq 10 M_{\mathrm{BH}}$ and $a_0\simeq100 M_{\mathrm{BH}}$). The solid white and dashed red curves mark the DM and Kerr ISCO boundaries, respectively.}
    \label{fig:DM_GRT}
\end{figure*}

We now explore how a dense DM halo affects the reflection spectrum. In the top panel of Figure~\ref{fig:DM_GRT}, we compare the Fe\, K$\alpha$ lines generated in the DM metric (solid lines) against those from the standard Kerr metric (dash-dotted lines). For varying BH spins at a fixed observer inclination of $i=50^\circ$ (top left), the presence of the DM halo shifts the line profiles to lower energies compared to their vacuum Kerr equivalents. 

The spin sequence in the upper-left panel of Figure~\ref{fig:DM_GRT} shows two main features. First, the selected DM profiles are displaced toward lower energies relative to their mass- and spin-matched Kerr counterparts. We explain the origin of this displacement in the next subsection. Second, the profiles vary less strongly with spin in the DM spacetime, consistent with the weaker spin dependence of the ISCO radius for these halo solutions. The inclination sequence at fixed $j=0.528$ shows that the redward displacement is also present at the sampled viewing angles. Increasing the inclination broadens the line through the line-of-sight Doppler shift and moves the blue horn to higher energy, but the DM and Kerr profiles remain distinguishable in these examples.

In the lower panels of Figure~\ref{fig:DM_GRT}, we compare accretion-disk maps in the Kerr and DM metrics at observer inclinations $i=10^\circ$ and $80^\circ$. At low inclination, the ratio is nearly azimuth-independent and $g_{\rm DM}/g_K<1$ over most of the disk, reflecting the additional gravitational time dilation. At high inclination, the line-of-sight factor introduces a stronger azimuthal asymmetry: it reinforces the redshift on the receding side but partly offsets it on the approaching side. For one of the displayed configurations, this compensation produces a localized region with $g_{\rm DM}/g_K>1$. The feature is therefore a differential change between the two spacetimes, not simply the ordinary Doppler blueshift present in either disk separately.

\subsection{Extra Redshift from a Dark Matter Halo}\label{sec:redshift}

To understand the physical origin of the line shifts, we decompose $g_{\rm DM}/g_K$ into contributions associated with the lapse, kinematic time dilation, and trajectory-dependent geometrical-Doppler effects. We compare radial quantities at the same circumferential radius, $R=\sqrt{g_{\phi\phi}}$.

\begin{figure*}[ht]
    \centering
    \includegraphics[width=\linewidth]{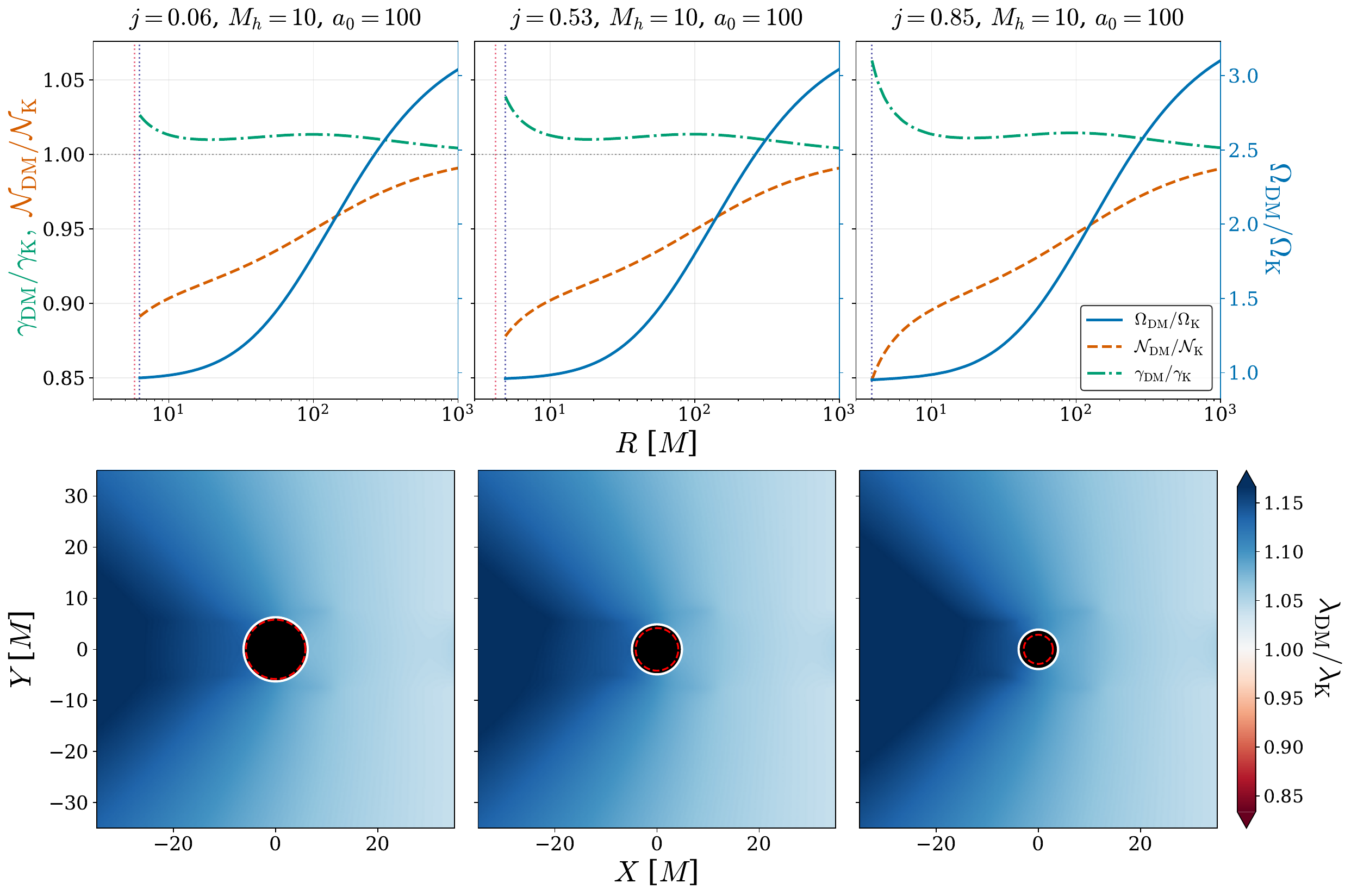}
    \caption{Decomposition of the quantities entering the $g$-factor ratio between the DM and Kerr spacetimes. Columns show
    $j\simeq0.06$, $0.53$, and $0.85$ from left to right, respectively, with $M_h\simeq10\,M_{\rm BH}$ and $a_0\simeq100\,M_{\rm BH}$. In the upper panels, the orange dashed and green dot-dashed curves show $\lapse_{\rm DM}/\lapse_K$ and $\gamma_{\rm DM}/\gamma_K$ on the left axis, while the blue solid curve shows $\Omega_{\rm DM}/\Omega_K$ on the right axis. These radial quantities are independent of the observer inclination. The lower panels show $\lambda_{\rm DM}/\lambda_K$ over the equatorial disk for rays received at $i=60^\circ$, using the disk-plane coordinates $X=R\cos\phi$ and $Y=R\sin\phi$. Solid white and dashed red circles denote the DM and Kerr ISCOs, respectively.}
    \label{fig:9_panel}
\end{figure*}

For gas following a circular equatorial orbit, the time component of the four-velocity can be written in terms of quantities measured by a ZAMO as
\begin{equation}\label{eq:ut}
    u_e^t=\frac{\gamma}{\lapse},
\end{equation}
where $\gamma=(1-v^2)^{-1/2}$ is the Lorentz factor associated with the total orbital speed $v$ measured by the ZAMO. The lapse function is
\begin{equation}
    \lapse=(-g^{tt})^{-1/2}
    =\sqrt{R^2\omega^2-g_{tt}},
\end{equation}
where $\omega=-g_{t\phi}/g_{\phi\phi}$ is the frame-dragging angular velocity. The coordinate angular velocity of the gas is $\Omega=d\phi/dt$ \citep{Bardeen72, Cunningham73}.

For a photon with conserved impact parameter $\lambda=\mathcal{L}_z/\mathcal{E}$, the observed-to-emitted energy ratio is
\begin{equation}
    g=\frac{1}{u_e^t(1-\lambda\Omega)}
     =\frac{\lapse}{\gamma(1-\lambda\Omega)}.
\end{equation}

The ratio between the energy shifts in the DM and Kerr spacetimes can
therefore be written exactly as
\begin{equation}\label{eq:gfactor3}
    \frac{g_{\rm DM}}{g_K}
    =
    \underbrace{
    \frac{\lapse_{\rm DM}}{\lapse_K}
    }_{\mathcal{G}_{\rm lapse}}
    \underbrace{
    \frac{\gamma_K}{\gamma_{\rm DM}}
    }_{\mathcal{G}_{\rm td}}
    \underbrace{
    \frac{1-\lambda_K\Omega_K}
         {1-\lambda_{\rm DM}\Omega_{\rm DM}}
    }_{\mathcal{G}_{\rm gd}} .
\end{equation}
Here, $\mathcal{G}_{\rm lapse}$ describes the difference in gravitational clock rates, while $\mathcal{G}_{\rm td}$ describes the difference in kinematic time dilation of the orbiting gas relative to the local ZAMO. The factor $\mathcal{G}_{\rm gd}$ is a trajectory-dependent geometrical-Doppler factor. It combines the orbital motion through $\Omega$ with the photon trajectory through $\lambda$. The first two factors depend only on the emission radius, whereas $\mathcal{G}_{\rm gd}$ also depends on disk azimuth and observer inclination.
\begin{figure*}
    \centering
    \includegraphics[width=0.46\linewidth]{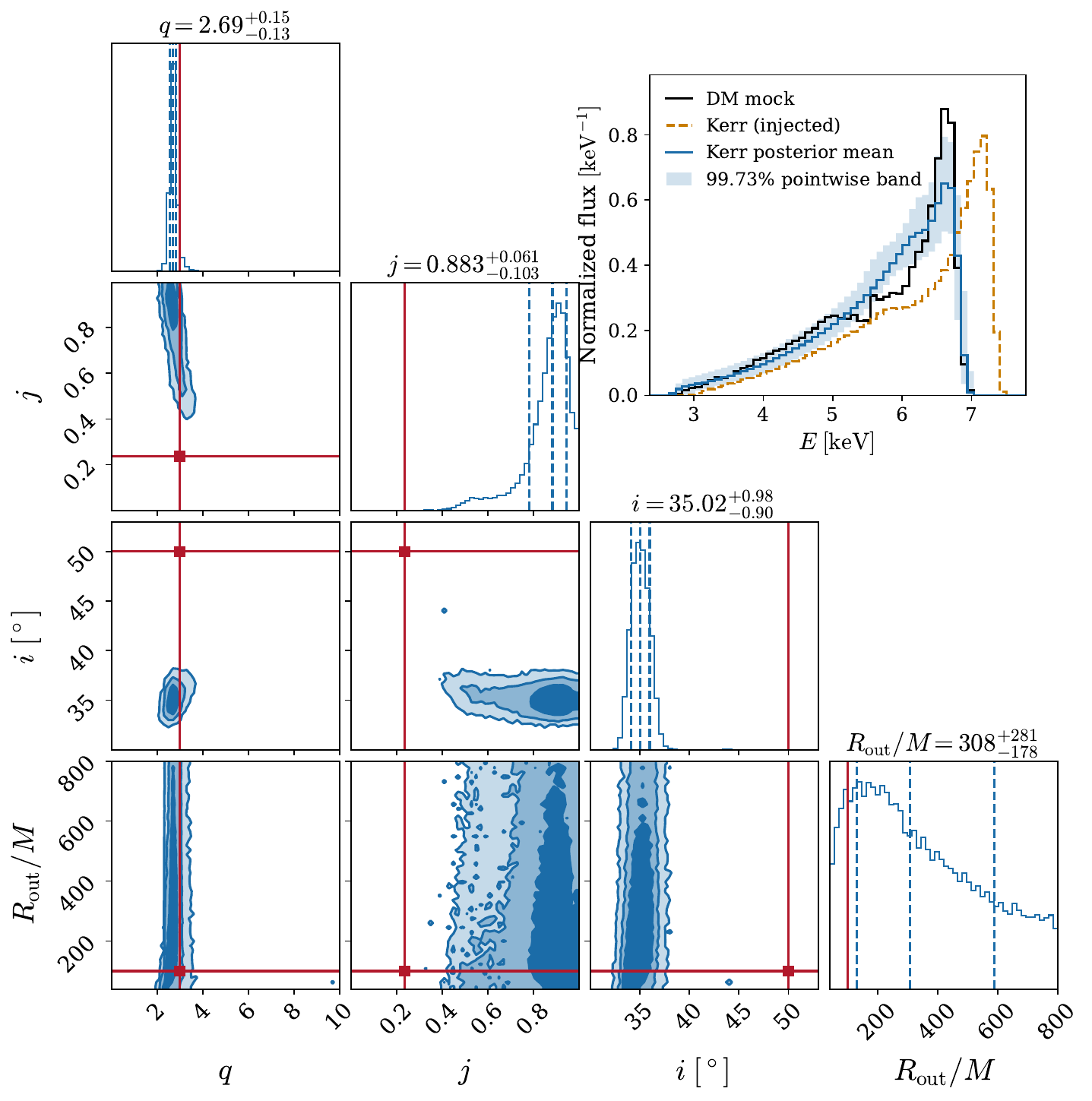}
    \includegraphics[width=0.46\linewidth]{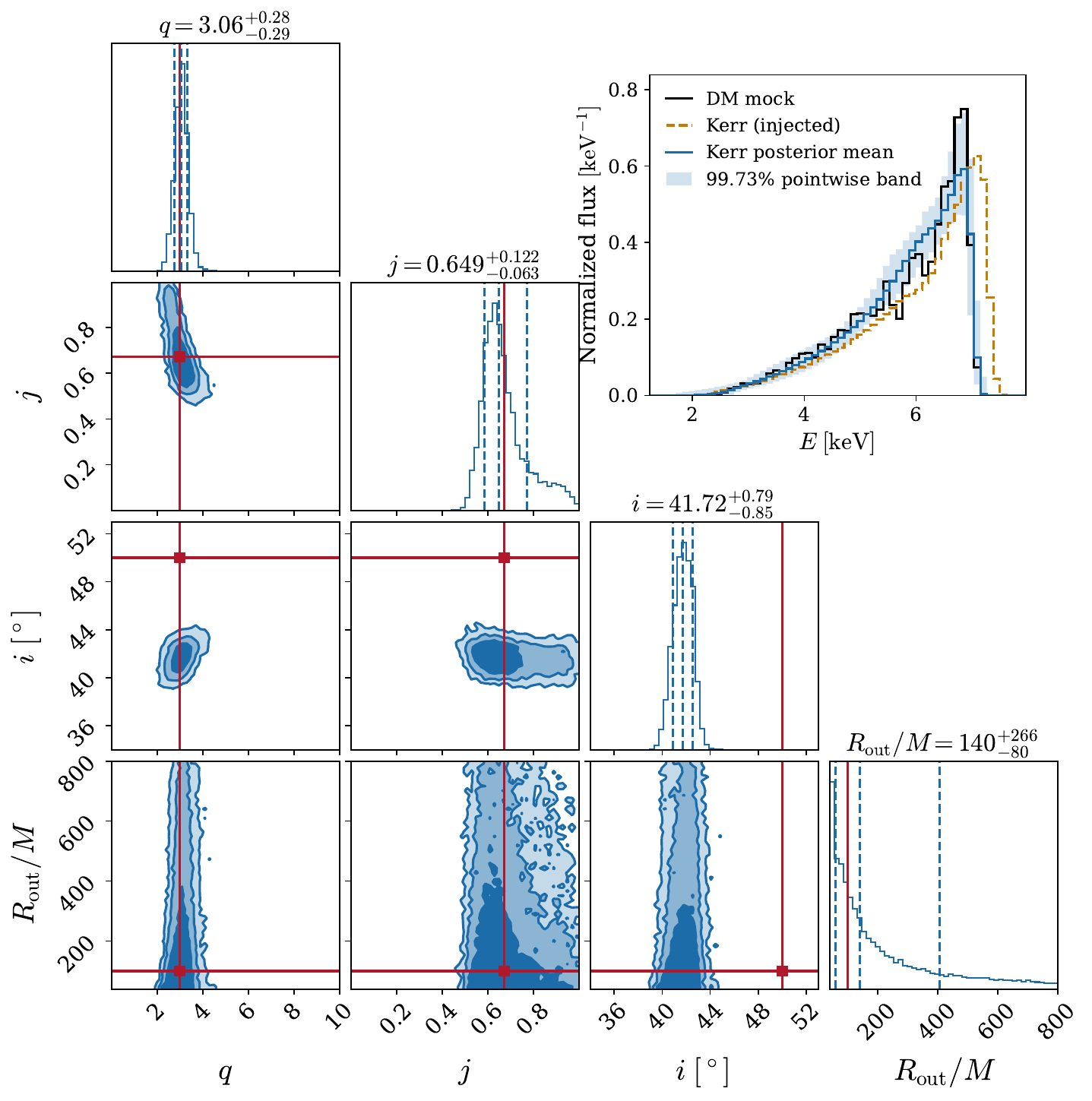}

    \caption{Illustrative parameter posteriors from fitting Kerr line shapes to DM-modified Fe\,K$\alpha$ lines with injected spins $j=0.236$ (left) and $0.672$ (right). Both mocks have $q=3$, $i=50^\circ$ and $R_{\rm out}=100M_{\rm BH}$; their halo compactness are approximately $0.1$ and $0.05$, respectively. Blue contours enclose $68$, $95$ and $99.73$ per cent of the marginal posterior for the finite-energy-resolution test, using $N_{\rm line}=300$ and the priors in Section~3.3. Red guides mark injected parameters. The upper-right insets compare the normalized DM mock line (black), the Kerr line at the injected parameters (orange dashed), and the posterior mean Kerr line (blue). All three inset curves use the same constant Gaussian response with FWHM $137\,\mathrm{eV}$ as the posterior calculation. The mean line is averaged over posterior draws and is not a single best-fit model. The light-blue band gives the pointwise central $99.73$ per cent range of posterior Kerr line shapes; it is not a simultaneous band or a data-error interval. This line-only response test contains no continuum and is not a full NICER observation. The full allowed spin interval is displayed.}
    \label{fig:MCMC}
\end{figure*}

The orange dashed curves in Figure~\ref{fig:9_panel} show that the lapse function ratio, $\lapse_{\rm DM}/\lapse_K<1$, over the displayed radial interval and approaches unity asymptotically. The lapse specifies the rate at which the proper time of a local ZAMO advances relative to the asymptotically normalized time coordinate. At a fixed circumferential radius, the halo lowers the lapse relative to Kerr, and the smaller inner-disk lapse produces a stronger gravitational redshift between the emitter and the distant observer. Because the lapse ratio enters Equation~\eqref{eq:gfactor3} directly, it contributes an azimuth-independent redshift and provides the largest contribution to the overall displacement of the selected line profiles.

The blue curves show that the response of $\Omega$ is radius-dependent. The angular velocity of a circular geodesic is determined by radial derivatives of the metric. In the inner region, the halo potential varies relatively slowly across an orbit but changes the normalization of coordinate time relative to infinity, tending to reduce $\Omega_{\rm DM}/\Omega_K$. Farther out, the halo contribution to the radial metric gradients becomes increasingly important relative to that of the BH and raises the angular velocity required for circular motion. The competition between these effects makes $\Omega_{\rm DM}/\Omega_K$ increase with radius and eventually exceed unity for the displayed solutions. The crossing radius identifies where the two effects balance and therefore depends on the mass and radial distribution of the halo.

The green curves show that $\gamma_{\rm DM}/\gamma_K>1$ throughout the displayed radial interval. The orbiting gas therefore experiences slightly stronger kinematic time dilation relative to the local ZAMO in the DM spacetime. The ratio varies weakly and approaches unity outwards. Its behavior follows from the local orbital speed $v=R(\Omega-\omega)/\lapse$, which depends jointly on the lapse, orbital angular velocity, and frame dragging. We discuss this relation further in~\ref{appendix_a}. Since
$\gamma_K/\gamma_{\rm DM}<1$, the time-dilation factor
$\mathcal{G}_{\rm td}$ contributes a modest redshift. Equivalently,
\begin{equation}
    \frac{u^t_{\rm DM}}{u^t_K}
    =
    \frac{\gamma_{\rm DM}}{\gamma_K}
    \frac{\lapse_K}{\lapse_{\rm DM}}
    >1,
\end{equation}
so the trajectory-independent contribution $1/u^t$ is smaller in the halo spacetime.

The geometrical-Doppler factor describes how the photon trajectory and orbital motion jointly affect the observed energy. The halo modifies the lensing map and therefore changes the value of $\lambda$ associated with a ray connecting a given disk element to the observer. At an asymptotic screen, $\lambda=-\alpha\sin i$ for the adopted screen convention, where $\alpha$ is the image-plane coordinate perpendicular to the projected spin axis \citep{Bardeen72, Cunningham73}. Rays reaching the same disk element in the two spacetimes can consequently originate from different screen coordinates and have different values of $\lambda$. The energy shift depends on the combination $\lambda\Omega$ through $(1-\lambda\Omega)^{-1}$.

On the approaching side of the disk, the geometrical-Doppler factor can partly compensate for the lapse and time-dilation suppression. At high inclination, this compensation can produce localized regions with $g_{\rm DM}/g_K>1$. On the receding side, the factor strengthens the redshift. It therefore changes the relative positions and amplitudes of the Doppler horns and introduces an inclination-dependent asymmetry into the line profile.

Combining the three factors in Equation~\eqref{eq:gfactor3}, the lapse suppression and enhanced kinematic time dilation provide an azimuth-independent redshift, while the geometrical-Doppler factor redistributes the energy shift between the approaching and receding sides of the disk. For the configurations shown here, the combined effect displaces the line towards lower energies and modifies its Doppler-broadened shape.

\subsection{Potential Influences on Parameter Estimation of Dark Matter Embedded Black Holes}

To explore whether a DM-modified line can resemble emission from a Kerr spacetime, we take simulated Fe\, K$\alpha$ profiles as mock lines and fit them with the Kerr model \texttt{relline} \citep{Dauser10}. We allow the emissivity index $q$, dimensionless spin $j$, observer inclination $i$, and outer disk radius $R_{\rm out}/M$ to vary. This retains a power-law emissivity without assuming a specific coronal geometry.

We bin each simulated line into 50 energy channels and adopt an effective line photon count of $N_{\rm line}=300$ for MCMC fits with \texttt{emcee} \citep{ForemanMackey13} using 16 walkers and consider the sampling as adequate when the retained chain length exceeds 50 times the maximum integrated autocorrelation time, $N_{\rm post}>50\tau_{\max}$. We visualize the posterior samples using \texttt{corner} \citep{ForemanMackey16}.

Figure~\ref{fig:MCMC} shows that Kerr models can reproduce the broad morphology of the DM-modified lines with different inferred parameters. For the injected $j=0.236$ case, we recover a median Kerr spin of $j=0.883$, illustrating how a higher spin can imitate part of the halo-induced redshift. For $j=0.672$, whose halo has a characteristic density scale $M_h/a_0^3$ approximately eight times lower, the inferred spin remains compatible with the injected value, although its posterior is asymmetric with a long tail towards high spin. The inferred inclinations are lower than the injected $50^\circ$ in both examples, with medians $35.02^\circ$ and $41.72^\circ$, respectively. The emissivity indices remain close to the injected $q=3$, with medians $q=2.69$ and $3.06$, respectively. The outer-radius posteriors are broad, with medians $R_{\rm out}/M\simeq308$ and $140$, respectively, compared with the injected value of $100$; the larger displacement occurs for the more compact halo. These results illustrate the possibility of spin overestimation in some configurations and motivate exploring whether reflection spectroscopy can help constrain halo compactness. The posterior widths remain illustrative rather than observational uncertainties; we examine response-folded spectra in Section~\ref{sec:xspec}.

\subsection{Simulated NICER and NuSTAR Observations}\label{sec:xspec}

\begin{figure}
    \centering
    \includegraphics[width=\linewidth]{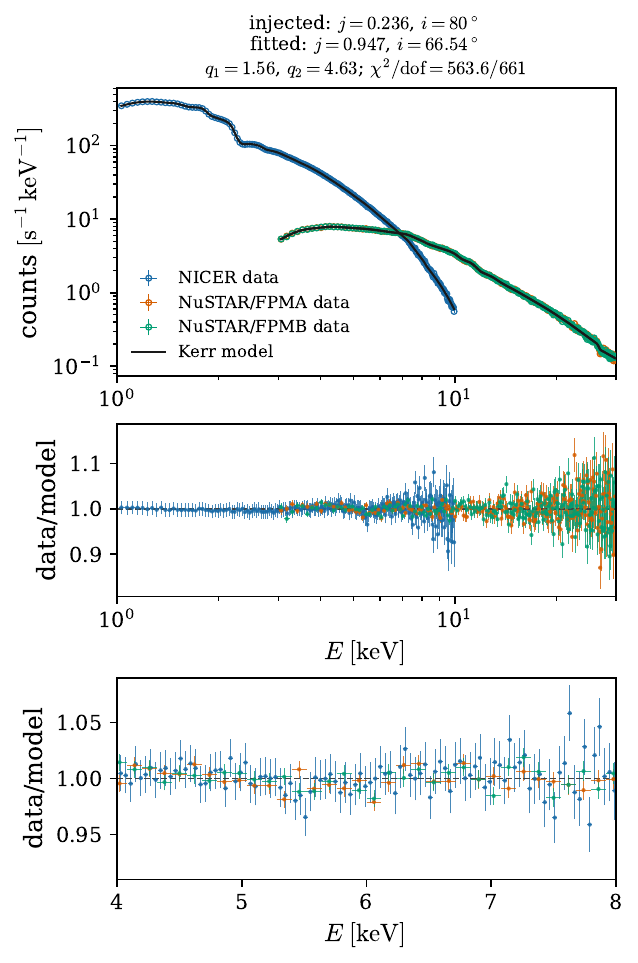}
    \caption{Simulated NICER and NuSTAR spectra for a DM-embedded BH with injected $j=0.236$ and $i=80^\circ$, fitted with the selected Kerr model. Top: count spectra, with open coloured markers showing the simulated data and black curves showing the response-folded model. Middle: data/model ratios over $1$--$10\,\mathrm{keV}$ for NICER and $3$--$30\,\mathrm{keV}$ for NuSTAR. Bottom: an expanded view of the Fe-line band. Blue, orange and green denote NICER, NuSTAR FPMA and NuSTAR FPMB. The selected fit has $j=0.947$ and $i=66.54^\circ$, with $R_{\rm out}=400M_{\rm BH}$ fixed. Parameters and nominal $90$ per cent uncertainties are given in Table~\ref{tab:fits}.}
    \label{fig:residuals}
\end{figure}

\begin{figure}
    \centering
    \includegraphics[width=\linewidth]{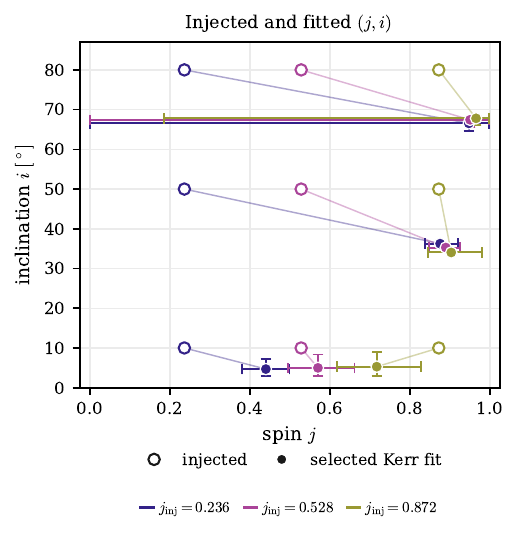}
    \caption{Comparison of injected and selected Kerr spin and inclination for the nine configurations in Table~\ref{tab:fits}. Open circles denote injected values and filled circles show selected fits. Indigo, magenta and olive denote injected spins $j=0.236$, $0.528$ and $0.872$, respectively; lines connect corresponding pairs. Horizontal and vertical bars give nominal $90$ per cent one-parameter profile ranges for spin and inclination. The outer disk radius is fixed at $R_{\rm out}=400M_{\rm BH}$ in all nine fits.}
    \label{fig:spins}
\end{figure}

\begin{figure*}
    \centering
    \includegraphics[width=\linewidth]{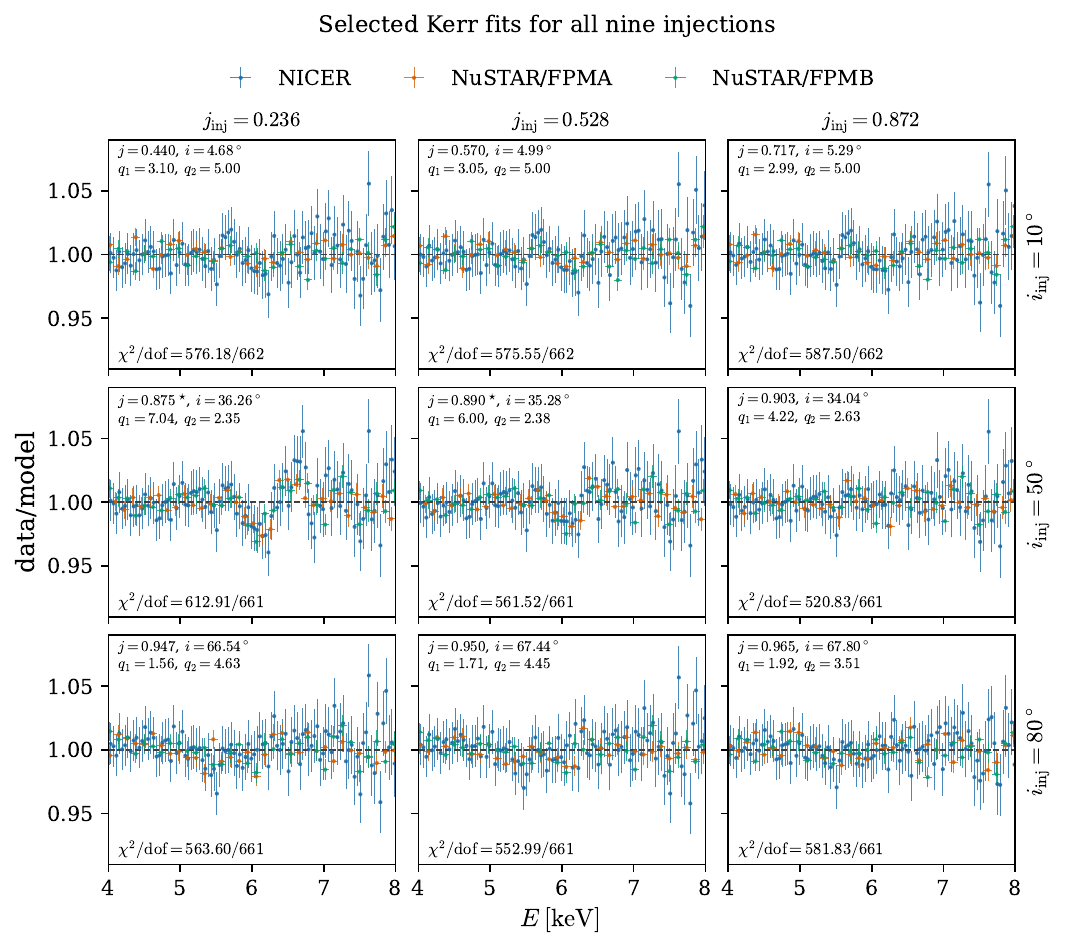}
    \caption{Fe-band data/model ratios for the nine selected Kerr fits in Table~\ref{tab:fits}. Columns correspond to injected spin and rows to injected inclination. Blue, orange and green points show NICER, NuSTAR FPMA and NuSTAR FPMB, respectively. Each panel gives the selected spin, inclination, broken power-law emissivity indices and fit statistic. Stars identify the selected high-spin local solutions in the first two $50^\circ$ cases. Fixed parameters and nominal $90$ per cent parameter ranges are listed in Table~\ref{tab:fits}.}
    \label{fig:grid}
\end{figure*}

\begin{figure*}
    \centering
    \includegraphics[width=0.94\linewidth]{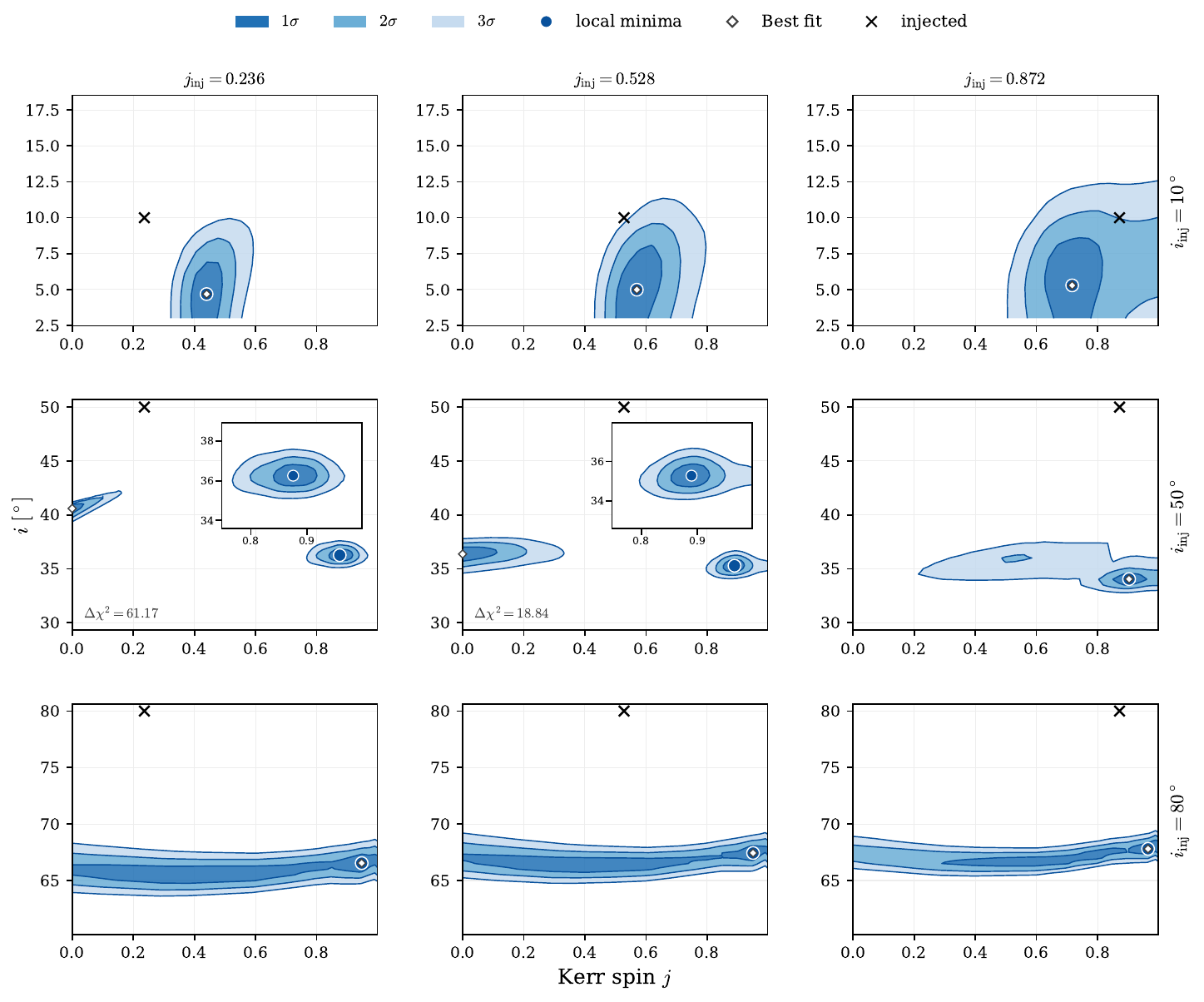}
    \caption{Spin--inclination profile-$\chi^2$ maps for the nine configurations. Columns and rows identify injected spin and inclination. Dark, medium and light blue show nominal joint $1$, $2$ and $3\sigma$ regions for two parameters, using $\Delta\chi^2=2.30$, $6.18$ and $11.83$ relative to the lowest statistic found for each configuration. Crosses mark injections, blue circles mark selected solutions, and white diamonds mark reference best fits. Additional high-spin islands and insets in the first two $50^\circ$ panels show local-branch diagnostics using the same increments relative to each selected local minimum. Their annotated offsets are measured from the configuration references. Insets preserve the relative scaling of the axes. All fits and maps fix $R_{\rm out}=400M_{\rm BH}$, and the $10^\circ$ cases also fix $q_2=5$, matching Table~\ref{tab:fits}.}
    \label{fig:post}
\end{figure*}

To assess how neglecting a compact DM halo could bias parameter estimates from X-ray observations, we simulate spectra for the Neutron star Interior Composition Explorer (NICER), a soft X-ray instrument mounted on the International Space Station, and the Nuclear Spectroscopic Telescope Array (NuSTAR), a focusing hard X-ray observatory. We then fit these spectra with standard Kerr models. We use XSPEC ~\citep{Arnaud96} to generate mock spectra for the nine combinations of spin ($j=0.236$, $0.528$ and $0.872$) and inclination ($i=10^\circ$, $50^\circ$ and $80^\circ$). The halos have $M_h\simeq10M_{\rm BH}$ and $a_0\simeq100M_{\rm BH}$. Each mock spectrum combines the DM-modified iron line with a smooth power-law continuum representing the underlying broad-band emission. We also account for absorption by foreground gas in the Milky Way. In XSPEC, these ingredients are combined as \texttt{TBabs*(powerlaw+atable)}: \texttt{TBabs} describes the absorption, \texttt{powerlaw} the continuum, and \texttt{atable} supplies the precomputed ray-traced line profile. For the line-emitting disk, we assume an emissivity proportional to $R_{\rm circ}^{-3}$, which weights the emission towards the inner disk, and an outer radius of $R_{\rm out}=100M_{\rm BH}$. The continuum slope is specified by a photon index $\Gamma=1.7$. The foreground absorption is set by a hydrogen column density $n_{\rm H}=6\times10^{21}\,{\rm cm}^{-2}$, which measures the amount of absorbing gas along the line of sight. We set the line strength relative to the continuum through an equivalent width of $250\,\mathrm{eV}$. Finally, we normalize the spectrum to an absorbed $2$--$10\,\mathrm{keV}$ flux of $2.4\times10^{-9}\,{\rm erg\,cm^{-2}\,s^{-1}}$ and simulate an exposure of $20\,\mathrm{ks}$ per detector. The simulations include instrumental responses, backgrounds, and Poisson counting noise. We simulate NICER and NuSTAR spectra using the \texttt{fakeit} command in XSPEC, following the procedure described in the XSPEC manual. We fold the input model through the corresponding instrumental responses and generate Poisson realizations of the expected counts for the adopted exposure times. We then group the simulated spectra to have at least $30$ total counts per bin.

For each simulated configuration, we jointly fit the NICER and NuSTAR spectra with a power-law continuum and a relativistically broadened iron line computed in the Kerr metric, both subject to foreground absorption. In XSPEC, this model is expressed as \texttt{constant*\allowbreak TBabs*(\allowbreak powerlaw+\allowbreak relline)}. We fix the rest-frame energy of the neutral Fe\,K$\alpha$ line at $6.40\,\mathrm{keV}$ and the cosmological redshift at zero. The inner disk edge is fixed at the Kerr ISCO corresponding to each trial spin. We adopt a broken power-law radial emissivity profile, with independent inner and outer indices, $q_1$ and $q_2$, and a free break radius, $R_{\rm br}$. The spin, inclination, emissivity parameters, outer disk radius, absorbing column density, continuum photon index, and continuum and line normalizations are allowed to vary. All detectors share the same physical source parameters. The two NuSTAR detectors have independent cross-normalization constants relative to NICER, whose constant is fixed to unity. We fit the NICER spectra over $1$--$10\,\mathrm{keV}$ and the NuSTAR spectra over $3$--$30\,\mathrm{keV}$. We restrict the NuSTAR band to emphasize the iron-line region because the simulated spectra contain only the prescribed power-law continuum at higher energies. Including these higher energies would therefore increase the statistical weight of the continuum in the fit without providing additional information about the line profile.

Figures~\ref{fig:residuals}--\ref{fig:grid} illustrate selected Kerr fits to the simulated NICER and NuSTAR spectra, with reduced fit statistics spanning $\chi^2/{\rm d.o.f.}=0.788$--$0.928$. Table~\ref{tab:fits} reports the selected fits and their parameter uncertainties. Eight of the nine selected fits return spins above the injected values. In two configurations with an injected inclination of $50^\circ$, parameter degeneracies give rise to both low-spin and high-spin fit solutions. The lowest-statistic fits found peg at the adopted lower bound of $j=0$, whereas the selected higher-spin solutions correspond to local minima with spin estimates away from the parameter boundaries.

To examine these competing solutions, we construct the spin--inclination profile maps shown in Figure~\ref{fig:post} by fixing spin and inclination and refitting the remaining parameters in XSPEC \citep{Lampton1976, Avni1976, Jia2022}. We use multiple initial values for the remaining parameters to explore competing solutions at each grid point. The blue circles mark the selected Kerr fits and additional local minima. We define $\Delta\chi^2=\chi^2-\chi^2_{\rm min}$ relative to the lowest statistic found for each configuration. The nominal joint $1$, $2$ and $3\sigma$ thresholds for two parameters are $2.30$, $6.18$ and $11.83$, respectively. For the two configurations at $i=50^\circ$ discussed above, the selected higher-spin solutions lie $\Delta\chi^2=60.81$ and $14.62$ above the corresponding lowest-statistic fits. They therefore provide alternative local fit solutions but lie outside the nominal joint $3\sigma$ confidence regions. These confidence regions are conditional on the adopted error weighting and parameter bounds, and the plotted shading is restricted to the evaluated grids.

\begin{table*}[t]
\caption{Selected Kerr fits to the simulated NICER and NuSTAR spectra. The outer disk radius is fixed at $R_{\rm out}=400M_{\rm BH}$ in all nine fits. Errors are nominal $90$ per cent one-parameter profile ranges, calculated with $\Delta\chi^2=2.706$ while refitting the other free parameters. NICER uses $1$--$10\,\mathrm{keV}$ and NuSTAR uses $3$--$30\,\mathrm{keV}$, retaining the adopted NICER systematic-error vector and case-specific parameter bounds. Inclinations are in degrees and $R_{\rm br}$ is in units of $M_{\rm BH}$. Bold injected parameters identify the two selected high-spin local solutions, whose ranges use the corresponding local minima.}
\label{tab:fits}
\centering
\begin{threeparttable}
\footnotesize
\renewcommand{\arraystretch}{1.20}
\setlength{\tabcolsep}{3pt}
\begin{tabular}{cccccccc}
\hline
$j_{\rm inj}$ &
$i_{\rm inj}$ &
$j$ &
$i$ &
$q_1$ &
$q_2$ &
$R_{\rm br}$ &
$\chi^2/{\rm d.o.f.}$ \\
\hline
$0.236$ & $10$ &
$0.440^{+0.059}_{-0.060}$ &
$4.68^{+2.57}_{-1.68}$ &
$3.10^{+0.20}_{-0.26}$ &
$5.00^{\dagger}$ &
$11.1^{+1.0}_{-1.0}$ &
$576.18/662$ \\

$0.528$ & $10$ &
$0.570^{+0.091}_{-0.076}$ &
$4.99^{+3.37}_{-1.99}$ &
$3.05^{+0.22}_{-0.28}$ &
$5.00^{\dagger}$ &
$10.9^{+1.2}_{-1.2}$ &
$575.55/662$ \\

$0.872$ & $10$ &
$0.717^{+0.111}_{-0.099}$ &
$5.29^{+3.64}_{-2.29}$ &
$2.99^{+0.21}_{-0.30}$ &
$5.00^{\dagger}$ &
$10.4^{+1.2}_{-1.2}$ &
$587.50/662$ \\
\hline
$\mathbf{0.236}$ & $\mathbf{50}$ &
$0.875^{+0.046}_{-0.039}$ &
$36.26^{+0.62}_{-0.56}$ &
$7.04^{+1.92}_{-1.41}$ &
$2.35^{+0.10}_{-0.10}$ &
$4.4^{+0.4}_{-0.4}$ &
$612.91/661$ \\

$\mathbf{0.528}$ & $\mathbf{50}$ &
$0.890^{+0.036}_{-0.042}$ &
$35.28^{+0.65}_{-0.62}$ &
$6.00^{+0.00}_{-0.97}$ &
$2.38^{+0.11}_{-0.12}$ &
$4.6^{+0.7}_{-0.5}$ &
$561.52/661$ \\

$0.872$ & $50$ &
$0.903^{+0.077}_{-0.059}$ &
$34.04^{+0.58}_{-0.61}$ &
$4.22^{+1.78}_{-0.95}$ &
$2.63^{+0.18}_{-0.19}$ &
$5.1^{+2.3}_{-2.1}$ &
$520.83/661$ \\
\hline
$0.236$ & $80$ &
$0.947^{\text{\S}}$ &
$66.54^{+0.93}_{-1.92}$ &
$1.56^{+0.60}_{-0.20}$ &
$4.63^{+0.37}_{-0.95}$ &
$34.0^{+5.9}_{-7.4}$ &
$563.60/661$ \\

$0.528$ & $80$ &
$0.950^{\text{\S}}$ &
$67.44^{+0.73}_{-1.86}$ &
$1.71^{+0.65}_{-0.17}$ &
$4.45^{+0.55}_{-0.85}$ &
$35.1^{+15.9}_{-8.1}$ &
$552.99/661$ \\

$0.872$ & $80$ &
$0.965^{+0.033}_{-0.780}$ &
$67.80^{+0.83}_{-1.73}$ &
$1.92^{+0.57}_{-0.23}$ &
$3.51^{+1.03}_{-0.47}$ &
$29.6^{+14.4}_{-10.2}$ &
$581.83/661$ \\
\hline
\end{tabular}
\begin{tablenotes}[flushleft]
\footnotesize
\item[$\dagger$] The emissivity index $q_2$ pegged at its upper bound of $5$ in the original fits and is fixed here.
\item[\S] The spin is unconstrained over the adopted range $0\leq j\leq0.998$.
\end{tablenotes}
\end{threeparttable}
\end{table*}

\section{Discussion and Conclusion}\label{discussion}

To summarize, we investigated the observational signatures of a DM halo surrounding a spinning BH, focusing on its impact on X-ray reflection spectroscopy. We adopted the numerical spectral solutions of \citet{Fernandes25} and performed geodesic ray tracing with an implementation adapted from \texttt{PyHole} to simulate the relativistically broadened Fe\, K$\alpha$ line.

Fitting the DM-modified profiles with standard vacuum Kerr models shows that the broad line morphology can be reproduced by changing the Kerr parameters. The asymmetric line-only posteriors and the response-folded confidence regions further illustrate degeneracies between the effects of the environment and those of the assumed Kerr geometry in the configurations examined.

In the simulated NICER and NuSTAR observations, eight of the nine selected Kerr fits return spins higher than the injected values (Table~\ref{tab:fits}), with $\chi^2/{\rm d.o.f.}=0.788$--$0.927$. The fits have $662$ degrees of freedom for the $10^\circ$ injections and $661$ for the $50^\circ$ and $80^\circ$ injections. The configurations at $i=80^\circ$ yield very high spin estimates, with $j\simeq0.95$--$0.97$, including for intrinsically slowly spinning BHs. However, the spin uncertainties for these configurations are broad or unconstrained over the adopted parameter range, so their high-spin estimates do not establish tightly constrained spin overestimation. In addition, some selected fits correspond to local minima rather than the lowest fit statistics found. These results illustrate how halo-induced line distortions can be accommodated by Kerr fits with substantially different spin estimates. Their implications for observational spin measurements must nevertheless be assessed in light of the limitations of our modeling.

Our methodology has several limitations. We assumed a power-law emissivity profile and modeled the Fe\,K$\alpha$ line rather than the full reflection spectrum. Although our simulated observations include a coronal power-law continuum and instrumental effects, they do not include the complex reflection continuum, such as the Compton hump and soft X-ray features, whose detailed modeling requires accounting for elemental abundances and the ionization state of the disk. Our illustrative fits over the $1$--$30\,\mathrm{keV}$ band therefore cannot capture all spectral distortions induced by a halo. A more complete spectral model and a broader-band analysis are needed to determine whether the tendency toward higher selected spin estimates persists. Moreover, we considered only a limited set of halo metrics and line configurations. Our metric assumes a stationary equilibrium halo that does not interact dynamically with the accreting gas; in realistic environments, accretion of DM or dynamical friction could induce secular changes to the metric or alter the disk structure.

We retained these limitations primarily to maintain a manageable computational cost, as the parameter space is vast and the steep gradients introduced by the DM halo require very high numerical resolution. Beyond computational constraints, these idealizations serve a distinct theoretical purpose. Furthermore, the modified field equations governing these spacetimes are highly non-linear and can consist of hundreds or even thousands of independent terms. Restricting our study to a stationary equilibrium halo is a necessary first step, as solving the coupled partial differential equations even in this axisymmetric limit is highly challenging and requires specialized, arbitrary-precision numerical techniques~\cite{Fernandes23, Fernandes25}. Direct integration of numerical DM metric solvers into broadband reflection convolution models would address these limitations, allowing simultaneous fitting of continuum and reflection features over a wider energy range.

Looking forward, DM-parameterized reflection models that treat the halo mass and characteristic scale as free parameters could be applied to both archival and future observations. If degeneracies similar to those found here persist in broadband analyses, comparison between vacuum and environmental models could test whether the data constrain a halo contribution. Parameter degeneracies are already well recognized in X-ray reflection spectroscopy, and allowing for DM halos may further complicate parameter inference. Future work could assess whether combining reflection spectroscopy with additional observables, such as X-ray polarization, helps constrain the accretion geometry and reduce degeneracies between black-hole properties and environmental effects.

Gravitational-wave observations provide a complementary probe because environmental effects can accumulate over the long inspiral. These include interactions with gaseous accretion disks \citep{Duque_2026, Lui2025, Lui2026, Hegade2025, Hegade2025b}, collisionless DM spikes \citep{Cardoso22b, Karydas_2025, Karydas_2026, Vicente_2025}, and ultralight-boson environments, including superradiantly grown scalar clouds \citep{Dyson_2025, Vicente22}. Combining electromagnetic and gravitational-wave information may help separate intrinsic spin from environmental effects.

\section*{Acknowledgements}

We thank Pedro G. S. Fernandes for providing the metric data to validate our results. We thank Lixin Dai, Pedro G. S. Fernandes, Javier A. Garc\'{i}a, Adam Ingram, Mingjun Liu and Yue Zhao for the helpful discussions. Finally, we thank the anonymous referee for constructive comments and suggestions that helped improve this work. LL and ATO were supported by the Beijing Natural Science Foundation (No. IS25014). ATO acknowledges support from the National Natural Science Foundation of China (No. W2533010).

\section*{Software}
\texttt{PyHole} \citep{Cunha16},
          \texttt{XSPEC} \citep{Arnaud96}, \texttt{relxill} \citep{Garcia10, Dauser13, Garcia14},
          \texttt{emcee} \citep{ForemanMackey13},
          \texttt{corner} \citep{ForemanMackey16},
          \texttt{numpy}, \texttt{scipy}, \texttt{matplotlib}

\appendix

\setcounter{equation}{0}
\renewcommand{\theequation}{A\arabic{equation}}
\renewcommand{\theHequation}{A.\arabic{equation}}
\setcounter{figure}{0}
\renewcommand{\thefigure}{A\arabic{figure}}
\renewcommand{\theHfigure}{A.\arabic{figure}}

\section{Gravito-Electromagnetic Decomposition of Orbital Kinematics}\label{appendix_a}

\begin{figure*}
    \centering
    \includegraphics[width=\linewidth]{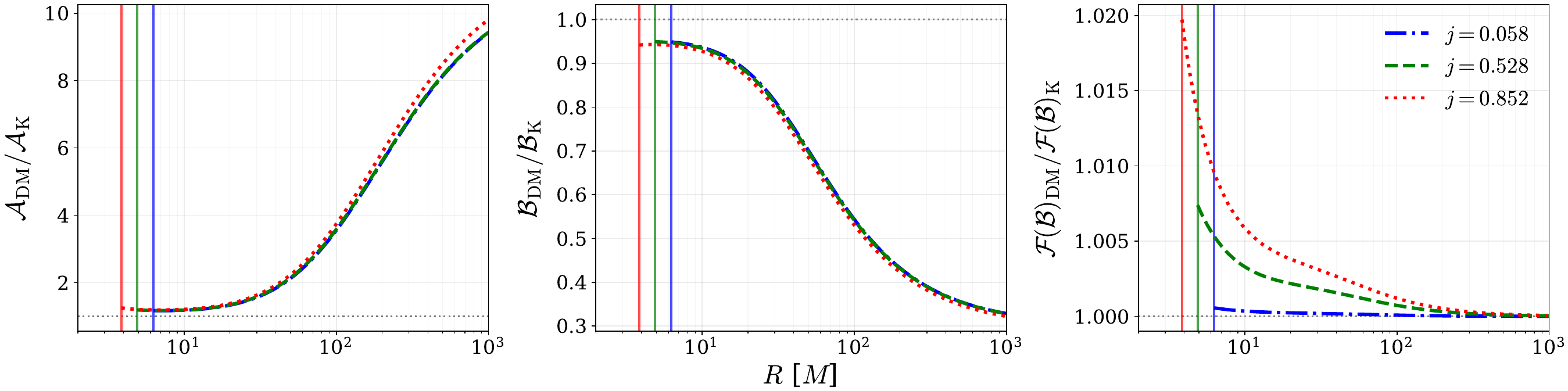}
    \caption{Radial profiles comparing the DM halo spacetime with the spin-matched Kerr metric. The panels show the ratios of the effective gravitoelectric quantity $\mathcal{A}$ (left), the effective gravitomagnetic quantity $\mathcal{B}$ (middle), and the kinematic function $\mathcal{F}(\mathcal{B})$ (right), plotted against the equatorial circumferential radius $R_{\rm circ}$. Curves correspond to $j\simeq0.06$, $0.53$, and $0.85$, with $M_h\simeq10\,M_{\rm BH}$ and $a_0\simeq100\,M_{\rm BH}$. Vertical lines denote the DM ISCO for each spin.}
    \label{fig:gravito_ratios}
\end{figure*}

To relate the Lorentz factor $\gamma$ to the local gravitational fields, we use a ZAMO-based gravito-electromagnetic (GEM) decomposition of prograde equatorial circular geodesics \citep{Bardeen72, Jantzen92, Gourgoulhon10, Paschalidis17}. In this description, the local orbital speed can be written as
\begin{equation}
v^2 = \mathcal{A}\mathcal{F}(\mathcal{B})^2,\quad \mathcal{F}(\mathcal{B}) = \sqrt{1+\mathcal{B}^2} + \mathcal{B},
\end{equation}
where 
\begin{equation}
    \mathcal{A} = \frac{d\ln \mathcal{N}}{d\ln R_{\rm circ}},\quad
    \mathcal{B} = \frac{R_{\rm circ}\omega'}{2\Omega_0},\quad
    \Omega_0=\sqrt{\frac{\mathcal{N}\mathcal{N}'}{R_{\rm circ}}},
\end{equation}
and a prime denotes $d/dR_{\rm circ}$. The quantity $\mathcal{A}$ measures the logarithmic radial gradient of the lapse, while $\mathcal{B}$ measures the frame-dragging gradient relative to the nonrotating circular-orbit scale $\Omega_0$. The function $\mathcal{F}(\mathcal{B})$ then describes how the gravitomagnetic term modifies the gravitoelectric contribution to the orbital speed and hence to $\gamma=(1-v^2)^{-1/2}$.

Figure~\ref{fig:gravito_ratios} shows that the principal radial change comes from $\mathcal{A}$. Near the BH, the logarithmic lapse gradient is still dominated by the central object and the DM-to-Kerr ratio remains comparatively close to unity. With increasing radius, the halo contributes a growing fraction of the radial gravitational field, so $\mathcal{A}_{\rm DM}/\mathcal{A}_K$ rises. The ratio can become large far outside the line-emitting region because the Kerr reference field continues to decline while the enclosed halo mass is still increasing.

The behavior of $\mathcal{B}$ is different. The halo is at rest relative to the ZAMO and has vanishing integrated angular momentum, so it does not add an independent rotating source comparable to the BH. It nevertheless changes the lapse gradient and hence the normalization $\Omega_0$, while also modifying the metric through which the BH frame dragging is transmitted. Consequently, $\mathcal{B}_{\rm DM}/\mathcal{B}_K$ decreases outwards for the displayed solutions. The corresponding ratio $\mathcal{F}(\mathcal{B})_{\rm DM}/\mathcal{F}(\mathcal{B})_K$ stays close to unity across the line-emitting disk. Thus, for these metrics, the modest enhancement of $\gamma$ is associated mainly with the change in the gravitoelectric lapse gradient, with the gravitomagnetic term providing a smaller spin-dependent correction. This decomposition explains why $\gamma_{\rm DM}/\gamma_K$ can remain above unity even where the coordinate-frequency ratio $\Omega_{\rm DM}/\Omega_K$ is below unity: $\Omega$ is defined relative to asymptotic coordinate time, whereas $\gamma$ is a local ZAMO quantity governed by both fields. 

\section*{Declaration of generative AI and AI-assisted technologies in the manuscript preparation process}
This work uses large language models (LLMs), specifically ChatGPT-6 Astra, 5.6/6.1 Sol, Claude Opus 5.5, DeepSeek V4/4.1 and Gemini 3.1 for coding, optimization, documentation, manuscript structuring and drafting. The authors maintained strict oversight over all LLM applications and are responsible for the scientific content, interpretation, and conclusions of this work.

\bibliographystyle{cas-model2-names}
\bibliography{sample701}

\end{document}